\documentclass[]{raa}            

\usepackage{graphicx,times}             
\usepackage{natbib}
\usepackage{amssymb,amsmath}
\bibpunct{(}{)}{;}{a}{}{,}

\usepackage[a4paper=true,dvipdfm=true,pagebackref=true]{hyperref}
\hypersetup{colorlinks = true, linkcolor = green, anchorcolor = red, citecolor = blue, filecolor = red, pagecolor = red, urlcolor = red}

\begin{document}

   \title{Spectral study of V565 Mon: Probable FU Ori-like or chemically peculiar star}

   \volnopage{Vol.0 (20xx) No.0, 000--000}
   \setcounter{page}{1}    

   \author{H. R. Andreasyan
      \inst{}
   }
   

   \institute{Byurakan Astrophysical Observatory, NAS RA,
             Byurakan, Aragatzotn Prov., 0213, Armenia; {\it hasmik.andreasyan@gmail.com}}
        
\vs\no
   {\small Received~~20xx month day; accepted~~20xx~~month day}

\abstract{ We present the detailed spectroscopic study of V565 Mon pre-main-sequence star, which is the illuminating star of Parsamian 17 cometary nebula. Observations were performed with 2.6 m telescope in Byurakan Astrophysical Observatory in 15 February 2018. Radial velocities and equivalent widths of the most prominent lines of V565 Mon are presented. We built the spectral energy distribution and estimated the main parameters of the star, for example obtained bolometric luminosity of V565 Mon is $L_{V565}\approx 130L_{\odot} $ . Considering all features of V565 Mon we came to conclusion that this young intermediate-mass star can belongs to some intermediate class between T Tau and HAeBe stars. Very unusual for a young star is the presence of strong absorption Ba II lines in the spectrum. Possible explanations on this issue are discussed. Hence, we think that V565 Mon is a unique example, which can help to understand some open questions in nucleosynthesis problem in young stars. 
}

   \authorrunning{H.R. Andreasyan }            
   \titlerunning{Spectroscopy of V565 Mon }  

   \maketitle


%
%
\section{Introduction}           
\label{sect:intro}

V565 Mon belongs to the little studied star formation region. As a variable star, it was discovered by \citet{Hoffmeister1968}. It is the illuminating star of the Parsamian 17 reflection cometary nebula, in which it is deeply embedded \citep{Pars1965, Cohen1974}. The observational data about this star are scarce. First description of the morphology of the nebula and the multi-band infrared (IR) observations of the central star, which revealed that V565 Mon is a prominent IR source, were given in the work of \citet{Cohen1974}. Only after 10 years this object was reobserved photometrically by \citet{Neckel1984}. The spectrum of the star was shortly described in the \citet{Herbig1988} catalog, where V565 Mon is listed as HBC 546. The spectral type was very roughly estimated as G, with H$\alpha$ emission line and very strong BaII lines in the red part. 
A next important step in the study of this star was the identification of V565 Mon with the IRAS 06556-0752 source in the catalog of \citet{Weintraub1990}. This confirmed that V565 Mon is extremely bright in middle-IR range as one can see in Fig. \ref{Fig:2}. 

P17 cometary nebula, also known as NGC 2313, PP67 and GN 06.55.6.01 \citep{Magakian2003}, is an object with high surface brightness in the dark cloud LDN 1653, which distance is 1060 $\sim1200$ kpc \citep{Kim2004, Maddalena1986, Hilton1995}. P17 has triangular shape and V565 Mon is located in its south-western angle. On the deep images with high spatial resolution one can note the traces of the second, opposite cone, which suggests the bipolar structure of the nebula.    
In the course of the search of Herbig-Haro (HH) objects a group HH 947 A/B were discovered \citep{Magakian2008} near the symmetry axis of P17. Logically, V565 Mon can be considered as a source of this flow.

\begin{figure}[ht!]
   \centering
   \includegraphics[width=10cm]{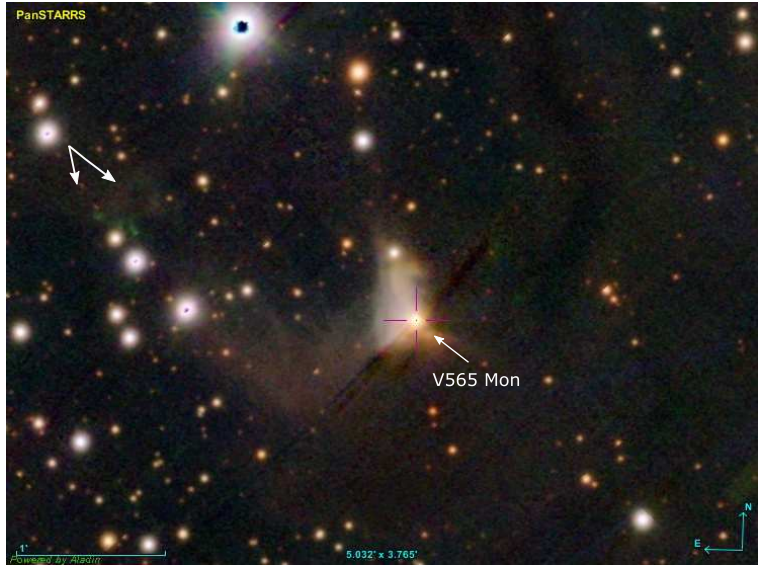}
   \caption{Color image of a field around V565 Mon from PanSTARRS DR1 survey (i, r, g filters). HH objects are pointed by arrows in the left corner. The dark diagonal stripe, visible across the nebula, is an artifact.}
   \label{Fig:1}
   \end{figure}

\begin{figure}[h!]
   \centering
   \includegraphics[width=10cm]{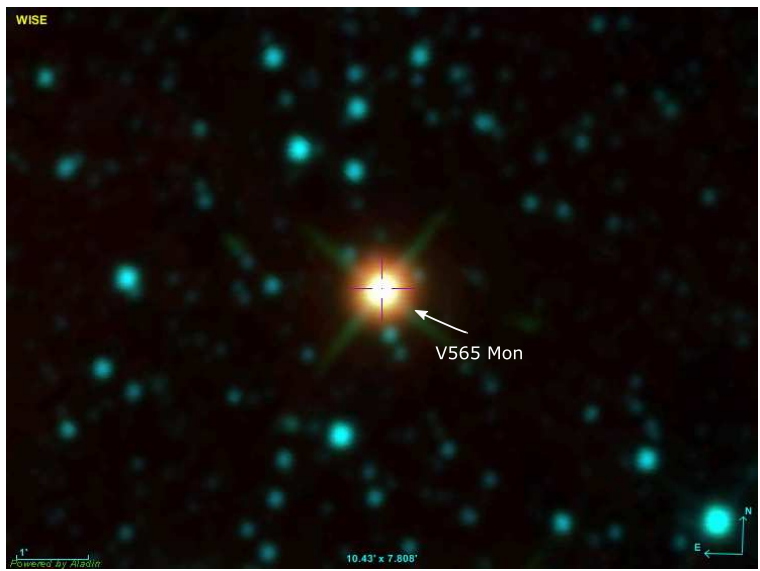}
   \caption{The same field from WISE survey (blue - 3.4 $\mu m$, green - 4.6 $\mu m$, red - 22 $\mu m$). V565 Mon is pointed by white arrow.}
   \label{Fig:2}
   \end{figure}

In the image taken from PanSTARRS survey one can see V565 Mon immersed in dust. Aforementioned HH group is also visible and pointed by arrows in the left corner of Fig. \ref{Fig:1}. For comparison in Fig. \ref{Fig:2} we present the same field taken from AllWISE survey coloured chart.

All of the above makes the V565 Mon star an interesting target for the further studies. Especially should be noted that we still do not have the good spectral data about this star. Thus, it was included in our program of spectral studies of selected PMS stars.

\section{Observations and Data Reduction}
\label{sect:Obs}

Observations were carried out on 2.6-m telescope of Byurakan Astrophysical Observatory in 15 February 2018. We used the SCORPIO spectral camera \citep{Scorpio} at prime focus of the telescope. As a detector the e2v CCD42-40 $2080 \times 2048$ CCD matrix was used, which works in imagery and long-slit modes. In the long-slit mode the width of the slit was $1.5^{\prime\prime}$ 
(with seeing about $2.5^{\prime\prime}$)  and the length was about $5^{\prime}$. As a dispersive element the volume phase holographic grating with 1800 g/mm was used providing the spectral resolution of about R=2500.
For the wavelength calibration the Ne+Ar lamp was used as a comparison spectra. Total exposure was 3600 s., which provides S/N ratio more than 100 in the final spectra after the processing and optimal extraction. 
Data reduction was done in usual way, using ESO-MIDAS program and appropriate packages. The stellar spectrum was extracted in $2^{\prime\prime}$ width zone, while background and night-sky emission lines were modeled along the all slit length. Our observations cover approximately 5800-6900 \AA \ wavelength range.

\section{Results}
\subsection{General description of the V565 Mon spectrum}
\label{sect:General}

\begin{figure}[h!]
   \centering
   \includegraphics[width=13cm]{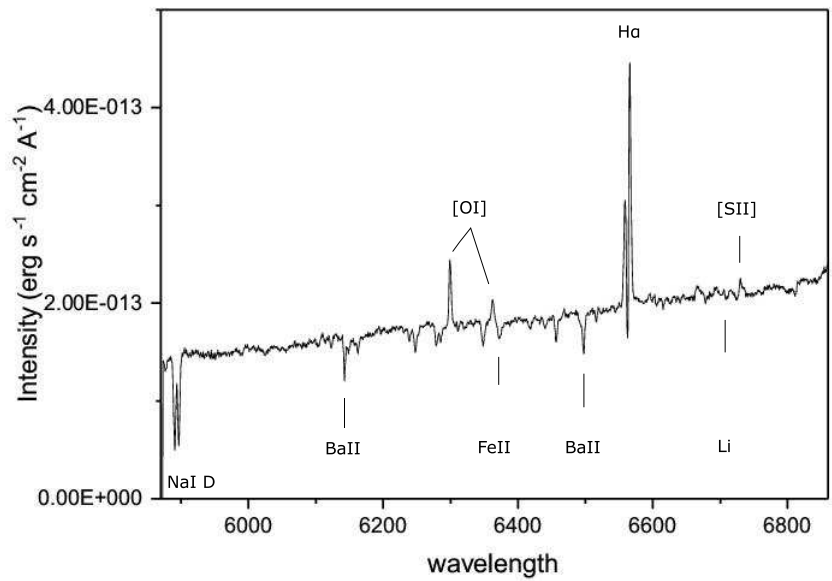}
   \caption{Spectrum of V565 Mon in absolute intensities}
   \label{Fig:3}
   \end{figure}

During our observations the visible brightness of V565 Mon was similar to the previously reported values. The reduced spectrum is presented in figure \ref{Fig:3}. One can see a red continuum with both emission and absorption lines superposed. Strong H$\alpha$ emission is divided by a relatively narrow (in comparison with the emission) absorption component, which is going below the continuum. Such double-peaked profile of H$\alpha$ line are typical for young active stars, although so strong central absorption is rare.

Among the absorption lines most prominent is a strong but quite narrow NaI D doublet. Besides, very noticeable are two $\lambda6141.71$ and $\lambda6496.89$ BaII absorption lines, which were already mentioned in \citep{Herbig1988} catalog. We also noticed weak 6707 LiI resonance line, which is well-known youth indicator for the PMS stars \citep{Herbig1964, Takeda2005, Lyubimkov2016}.  Also, a variety of FeII absorption lines exist, but none of FeI absorptions were detected.
Besides of H$\alpha$, only other emission lines detected are forbidden red doublets of [OI] and [SII]. Taking into account the presence of HH flow, associated with this star, one can assume that they belong to an outflowing envelope or jet near the star. The more or less accurate spectral type of V565 Mon still is difficult to estimate. In any case, the prominent NaI D lines, on the one hand, and well-developed FeII spectrum with absence of FeI lines, on the other hand, lead to the spectral range from late F to early G, confirming the Herbig's assumption. 

\subsection{Equivalent Widths and Radial Velocities}
\label{EqW}

We estimated the equivalent widths and heliocentric radial velocities of the most prominent absorption and emission lines and present them in Table \ref{tab1}.
In case of H$\alpha$ we list the separate measurements for emission and absorption components.

One can see that the negative radial velocities are observed only for forbidden lines, which confirms our assumption that these lines are related to HH outflow. Besides, low absolute values of their velocities point to the large angle of outflow to the line of sight. The absorption component of H$\alpha$ has near-zero velocity, and all other absorption lines, including NaI D, have positive velocities. In general the strength of emission lines is not high. Even for H$\alpha$ emission the total EW is  $\approx-7$  \r{A}.

\begin{table}[h!]
\caption{Equivalent Widths, and Heliocentric Radial Velocities of Selected Lines in V565 Mon Spectrum}
\label{tab1}  
\centering        
\begin{tabular}{c c c c}          
\hline\hline                        
Lines & \textit{EW, \r{A}} & \textit{$V_R$, $km s^{-1}$} & \textit{FWHM} \\ 
\hline                     
Na I D$_{2}$ (5889.96) & 2.10  & 36 & 1.94\\
Na I D$_{1}$ (5895.93) & 1.88 & 32 & 3.84\\ 
Ba II      (6141.71) &  0.83 & 62 & 2.62\\
\ [O I] (6300.31)  & $-$1.67 & $-$36 & 3.88\\
\ [O I] (6363.82)  & $-$0.58 & $-$51 & 4.65\\
Fe II (6456.39)  & 0.45 & 24 & 3.92 \\
Ba II      (6496.89) & 1.06 & 32 & 3.09\\
Fe II (6516.05) & 0.2 & 17 & 3.02 \\
H$\alpha$ (em) & $-$2.22 & $-$181 & 3.61\\
H$\alpha$ (abs) &  0.22  & $-$2 & 3.92\\
H$\alpha$ (em) &  $-$4.64  & 157 & 3.05\\
Li I (6707)    &  0.08  &  89 & 1.48\\
\ [S II] (6730.78)  &  $-$2.87  & $-$6 & 2.61\\

\hline
   
\end{tabular}
\end{table}

The mean radial velocity of V565 Mon, computed by six strongest absorption lines, is +34 $\pm$ 14 km s$^{-1}$. Considering our spectral resolution and the probable blending of several lines, such an error should be considered small enough. The line widths somewhat exceed the instrumental profile of our system, which confirms their photospheric origin.

\section{Discussion and Conclusions}
\label{sect:discussion}

First of all, we compared previously obtained estimations of V565 Mon distance and brightness with new data from Gaia DR2. The modern distance estimate for V565 Mon, based directly on the newly obtained Gaia parallaxes, is 1150 pc ($\pm\ 91$ pc), while the statistic estimation from the \citet{Bailer2018} catalogue is 1122 pc. This value is in accordance with the previous estimates of LDN 1653 distance (see Sec.\ref{sect:intro}). Brightness of the star also did not changed in a noticeable way.

\begin{figure}[ht!]
   \centering
   \includegraphics[width=9cm]{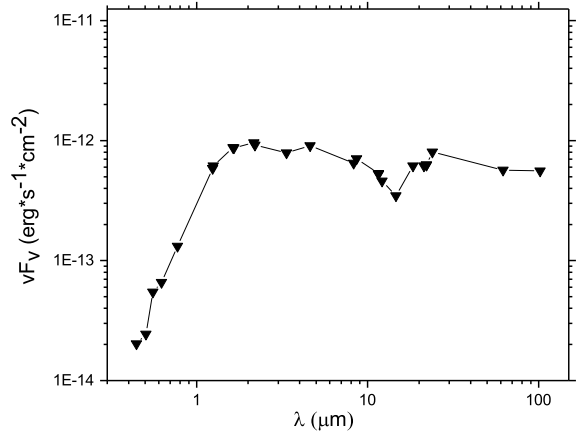}
   \caption{SED of V565 Mon}
   \label{Fig:4}
   \end{figure}

For the evaluation of main parameters of V565 Mon we obtained photometric data for this object with Vizier VO tools, including the photometry from IRAS and AKARI all-sky surveys \citep{Abrahamyan2015}, WISE survey \citep{Cutri2012}, also \citep{Fischer2016}, MSX catalog \citep{Egan2003}, 2MASS All-Sky catalog \citep{Cutri2003, Tian2017} and Gaia survey \citep{Gaia2018}. After making necessary transformations and calculations we got the spectral energy distribution (SED) of V565 Mon, which is presented in Fig. \ref{Fig:4}. Of course, one should keep in mind that at longer wavelenghts (e.g. in AKARI data) significant amount of emission can have extended origin; however, the SED, obtained by us, is more or less consistent with the star, surrounded by large mass of heated dust, very likely in the form of circumstellar disk.

The problem of the extinction value for V565 Mon also is important. In any case, since  the star is pretty well visible in the visual range, its extinction can not be too high. Gaia DR2 extinction value for V565 Mon is A$_G$ = 2.7. Using visible magnitude of V565 Mon  (V = 13.72) from \citet{Herbig1988} catalogue and distance of the star (estimated using Gaia DR2), we've got M$_{V}$ = 3.05. Comparing this value with standard M$_{V}$ for G0 type main sequence star \citep{Allen}, for extinction value one can obtain A$_V$ = 1.4. Of course, in case of V565 Mon this value must be higher, because V565 Mon is definitely located above the main sequence.    

As can be seen from SED, V565 Mon emits in mid and far IR range at least the same, or even higher amount of energy, as in the optical range. To estimate its total luminosity we tried several approaches.
By integrating the SED curve of V565 Mon we obtained $L_{V565}\approx 130L_{\odot} $ for its bolometric luminosity. But even this value is only its lower limit, since, judging by the SED, the star should emit significant energy up to the submillimeter range. By the equation, suggested for infrared sources observed by IRAS \citep{CRT}, we've got only 75$L_{\odot}$, which in our case can be considered only as an additional correction, to take into account the furthest  IR range. We also tried to approximate SED by Robitaille models \citep{Robitaille2007}, but were not satisfied by any of solutions, because they cannot well represent nearly flat far-IR side of the SED. The possible reason of this can be significant extended emission in far-infrared range, but the available data do not allow to check its existence.

Taking 130 L$_{\odot}$ as an initial estimate for V565 Mon bolometric luminosity and making assumption that the flux in all observed ranges is a result of the radiation of G0 type central star, we can use general equations like 
\begin{equation} 
M_{bol,\star}-M_{bol,\odot}=-2.5log_{10}(L_{\star}/L_{\odot})
\end{equation}
and 
\begin{equation}
\label{2}
m-M=5logd-5+A{_v}
\end{equation}

to convert the bolometric luminosity of V565 Mon to absolute magnitude. As a result we've got M$_{bol}$= $-$0.58. We have to add the bolometric correction, which for the G0 star is rather small \citep{Allen}, hence, M$_{V}$= $-$0.55. We know the observed visible magnitude of V565 Mon (m=13.72, see above), so from Eq. \ref{2} the value  of extinction for V565 Mon should be A$_{V}$ = 2.86. The similarity of this result with Gaia estimate is remarkable.
One can assume that A$_{V} \approx\ 3$ is reasonable estimate for V565 Mon extinction.

It is obvious, that V565 Mon indeed is located in LDN 1653 cloud, because our estimate of its radial velocity is in a perfect accordance with a  velocity (converted to heliocentric) of  $+29$  km s$^{-1}$, measured by $^{13}$CO line \citep{Kim2004}. 

Thus, we see that all main features of V565 Mon, described above, point to the PMS nature of this star. However, it is not too easy to classify this object, because its significant luminosity (higher than great majority of T Tau stars) suggests that it has intermediate mass and, consequently, belongs to HAeBe stars class. On the other hand, its spectral type is probably too late for such classification and its spectrum corresponds to T Tau stars. Nevertheless V565 Mon can belongs to some intermediate class between T Tau and HAeBe stars.   

But the most unusual feature of V565 Mon is the presence of two strong $\lambda6141.71$ and $\lambda6496.89$ BaII absorption lines in its spectrum. Although this fact was mentioned in the Herbig-Bell catalogue \citep{Herbig1988}, nowhere was stressed the peculiarity of barium overabundance in so young star. Meanwhile, though it is assumed that for low and intermediate ($1M_{\odot}\leq M\leq 3M_{\odot}$) mass stars barium emerges through \textbf{s-process}, recent studies have found the distinct excess of barium abundance in young stellar clusters \citep{D'Orazi2009, Desidera2011, Maiorca2011, D'Orazi2012, Mishenina2013}. The main trend, described in these works, is in favor of the anticorrelation between barium enrichment and the age of the cluster. In more recent work of \citet{Mishenina2015} new mechanisms of barium enrichment in young open clusters were suggested. Some possible explanation of Barium abundance are linking it to chromospheric activity, however significant correlation was not detected \citep{D'Orazi2017}. Despite all these various approaches, the processes producing barium in young stars still are hardly understandable.
 
The most obvious difference of V565 Mon with these studies listed above is that even the youngest open clusters described in these works have ages 10$^{8}$ years, while the age of V565 Mon cannot exceed several million years. On the other hand, equivalent width of BaII line (1.06) is even larger than upper limit of the curve of growth from the work of \citet{Mishenina2015} (see their fig.1). We screened the spectra of several hundreds PMS stars from the Cohen and Kuhi atlas \citep{CK} and did not found any object with similar strength of BaII lines. 

Besides, we also considered the possibility that V565 Mon star can be FU Ori-like object. Actually, this probability was the reason why we included this star in our observational program. The spectral type of V565 Mon and the scarcity of emission lines in its spectrum are in favor of FUor hypothesis. Also, it is well known that Ba II (especially $\lambda 6497$) lines existence is one of the most typical features in spectra of FUors.  However, there are several reasons against such suggestion, in particular: the photospheric absorption lines definitely are wider than in bona fide FUors and FU Ori-like stars; the H$\alpha$ profile does not show characteristic wide P Cyg type absorption; radial velocities of all absorptions, including even NaD doublet, are positive, i.e. indicate the absence of significant outflowing activity. Of course, one cannot exclude that V565 Mon can represent some non-typical case of FU Ori-like stars, or that its spectral characteristics are due to its orientation to the line of sight.

Anyway for the last 25 years little studies were carried out to reveal the nature of V565 Mon. Few previous spectral investigations give only a general view, without going into details. Scarce photometric data that we have in hand also give very generic and not complete picture, which do not let us to make a perception about the variability of V565 Mon. Meanwhile, as our spectral study revealed, this somewhat neglected object could be an important step in the understanding of the nucleosynthesis problems in young stars.
In this sense, the case of V565 Mon is unique and it deserves a detailed high resolution spectral study in the optical and near-IR range .

\begin{acknowledgements}

I wish to thank the referee for very helpful suggestions and comments.
I am very thankful to my supervisor Dr. Tigran Magakian for great support and advices, and also to our team-member Dr. Tigran Movsessian for kindly providing observational data on V565 Mon.
This work was supported by the RA MES State Committee of Science, in the frame of the research project number 18T-1C-329.
This work has made use of data from the European Space Agency (ESA) mission {\it Gaia} (\url{https://www.cosmos.esa.int/gaia}), processed by the {\it Gaia} Data Processing and Analysis Consortium (DPAC, \url{https://www.cosmos.esa.int/web/gaia/dpac/consortium}). Funding for the DPAC has been provided by national institutions, in particular the institutions participating in the {\it Gaia} Multilateral Agreement. This publication makes use of data products from the Wide-field Infrared Survey Explorer, which is a joint project of the University of California, Los Angeles, and the Jet Propulsion Laboratory/California Institute of Technology, funded by the National Aeronautics and Space Administration. We thank The Pan-STARRS1 Surveys (PS1) and the PS1 public science archive, which have been made possible under Grant No. NNX08AR22G issued through the Planetary Science Division of the NASA Science Mission Directorate and  the National Science Foundation Grant No. AST-1238877․

\end{acknowledgements}

\label{lastpage}


\begin{thebibliography}{}


  \bibitem[Abrahamyan et al.(2015)]{Abrahamyan2015}
  Abrahamyan, H., Mickaelian, A. \& Knyazyan, A., \ 2015, Astronomy and Computing, 10, 99

  \bibitem[Afanasiev \& Moiseev(2005)]{Scorpio} Afanasiev, V.L. \& Moiseev, A.V., \ 2005, Astronomy Letters, 31, 194 
  
  \bibitem[Allen(1975)]{Allen} Allen, C.~W.,  \ 1975,  Astrophysical Quantities (3rd ed.; Athlone: London)
   
  \bibitem[Bailer-Jones et al.(2018)]{Bailer2018} Bailer-Jones, C., Farnocchia D., Meech K. Brasser R., Micheli M., Chakrabarti S., Buie M. \& Hainaut O., \ 2018, \aj, 156, 58B

  \bibitem[Cohen(1974)]{Cohen1974} Cohen, M., \ 1974, \pasp, 86, 813

  \bibitem[Cohen \& Kuhi(1979)]{CK} Cohen M., Kuhi L.V., 1979, \apjs, 41, 743

  \bibitem[Connelley, Reipurth \& Tokunaga(2007)]{CRT}
Connelley, M., Reipurth, B. \& Tokunaga, A., \ 2007, \aj, 133, 1528
  
  \bibitem[Cutri et al.(2003)]{Cutri2003}
  Cutri, R.~M., Skrutskie, M.~F., van Dyk, S.,      Beichman, C.~A., Carpenter, J.~M., Chester, T., Cambresy, L., Evans, T., Fowler, J., Gizis, J., Howard, E., Huchra, J., Jarrett, T., Kopan, E.~L., Kirkpatrick, J.~D., Light, R.~M., Marsh, K.~A., McCallon, H., Schneider, S.,Stiening, R., Sykes, M. Weinberg, M., Wheaton, W.~A., Wheelock, S., \& Zacarias, N., \ 2003, Vizier Online Data Catalog, 2246, 0C
  
  \bibitem[Cutri et al.(2012)]{Cutri2012}
  Cutri, R. M. \& et al., \ 2012, VizieR Online Data Catalog, 2311, 0C
  

  \bibitem[Desidera et al.(2011)]{Desidera2011} Desidera S., Covino E., Messina S., D'Orazi V., Alcal{\'a} J.~M., Brugaletta E., Carson J., Lanzafame A.~C. \& Launhardt R., \ 2011, \aap, 529A, 54D
   
  \bibitem[D'Orazi et al.(2009)] {D'Orazi2009}  D'Orazi V., Magrini L., Randich S., Galli D., Busso M. \& Sestito P., \ 2009, \apj, 693L, 31D
    
  \bibitem[D'Orazi et al.(2012)] {D'Orazi2012}  D’Orazi V., Biazzo K., Desidera S., Covino E., Andrievsky S. M., \& Gratton R. G., \  2012, \mnras, 423, 2789
     
  \bibitem[D'Orazi et al.(2017)] {D'Orazi2017}  D'Orazi V., De Silva G.~M. \& Melo C.~F.~H., \  2017, \aap, 598A, 86D
  
  \bibitem[Egan et al.(2003)]{Egan2003} 
  Egan, M.~P., Price, S.~D., Kraemer, K.~E.,  Mizuno, D.~R., Carey, S.~J., Wright, C.~O., Engelke, C.~W., Cohen, M., \& Gugliotti, M.~G., \ 2003, VizieR Online Data Catalog, 5114, 0E
  
  \bibitem[Gaia Collaboration(2018)]{Gaia2018}
  Gaia Collaboration, Brown, A.~G.~A., Vallenari, A.,  Prusti, T., and de Bruijne, J.~H.~J., \ 2018, \aap, 616, 1
  
  \bibitem[Fischer et al.(2016)]{Fischer2016} Fischer, W.~J., Padgett, D.~L., Stapelfeldt, K.~L., Sewilo, M., \ 2016, \apj, 827, id.96
  
  \bibitem[Herbig(1964)]{Herbig1964} Herbig G.~H., \ 1964 \aj, 69Q, 141
   
  \bibitem[Herbig \& Bell(1988)]{Herbig1988} Herbig G.~H. \& Bell K.~R., \ 1988, Lick Obs. Bull. 1111
  
  \bibitem[Hilton \& Lahulla(1995)]{Hilton1995} Hilton, J. \& Lahulla, J. F., \ 1995 \aaps, 113, 325

  \bibitem[Hoffmeister(1968)] {Hoffmeister1968} Hoffmeister C., \ 1968, Astron. Nachrichten, 290, 277
  
  \bibitem[Kim et al.(2004)]{Kim2004} 
  Kim B. G., Kawamura A., Yonekura Y., Fukui Y., \
2004, PASJ, 56, 313

  \bibitem[Lyubimkov(2016)]{Lyubimkov2016}
  Lyubimkov L., \ 2016, Astrophysics, 59, 411

  \bibitem[Maddalena et al.(1986)]{Maddalena1986} Maddalena R.~J., Morris M., Moscowitz J. \& Thaddeus P., \ 1986 \apj, 303, 375

  \bibitem[Magakian(2003)]{Magakian2003} Magakian T. Yu., \ 2003, \aap, 399, 141

  \bibitem[Magakian et al.(2008)]{Magakian2008} Magakian T. Yu., Movsessian, T. A., Nikogossian E. G., \ 2008, \aap,  51, 7

  \bibitem[Maiorca et al.(2011)] {Maiorca2011}  Maiorca E., Randich S., Busso M., Magrini L. \& Palmerini S., \ 2011, \apj, 736, 120M
 
  \bibitem[Mishenina et al.(2013)] {Mishenina2013}  Mishenina T., Korotin S., Carraro G., Kovtyukh V.~V. \& Yegorova I.~A., \ 2013, \mnras, 433, 1436M
  
  \bibitem[Mishenina et al., (2015)] {Mishenina2015}  Mishenina T., Pignatari M., Carraro G., Kovtyukh V., Monaco L., Korotin S., Shereta E., Yegorova I. \& Herwig F., \ 2015, \mnras, 446, 3651M
      
  \bibitem[Neckel \& Staude(1984)]{Neckel1984} 
  Neckel T. \& Staude H.~J., \ 1984, \aap, 131, 200
  
  \bibitem[Parsamian(1965)]{Pars1965} Parsamian, E. S., \ 1965, \ IzArm, 18, 146
    
  \bibitem[Parsamian \& Petrossian(2019)]{Pars2019} Parsamian E.S. \& Petrossian V. M., \ 2019, \ CoBAO, 66, 70
  
  \bibitem[Robitaille et al. (2007)]{Robitaille2007} Robitaille, T. P., Whitney, B. A., Indebetouw, R., \& Wood, K., \ 2007, \apjs, 169, 328

  \bibitem[Takeda \& Kawanomoto(2005)]{Takeda2005} Takeda Y. \& Kawanomoto S., \ 2005 \pasj, 57, 45
  
  \bibitem[Tian, Hai-Jun et al. (2017)]{Tian2017}
  Tian, Hai-Jun, Gupta, P., Sesar, B., Rix, H., Martin, N., Liu, Ch., Goldman, B., Platais, I. Kudritzki, R. \& Waters, Ch., \ 2017 \apjs, 232, 4
  
  \bibitem[Weintraub(1990)]{Weintraub1990} Weintraub D. A., \ 1990 \apj  74, 575 

  \bibitem[Zacharias et al.(2013)]{UCAC}  Zacharias N., Finch C.T., Girard T.M., Henden A., Bartlett J.L., Monet D.G., Zacharias M.I., \ 2013 \aj, 145, 44 
 
  
   
     
       
     
    

   

\end{thebibliography}
\end{document}